\documentstyle[aps,epsfig]{revtex}

\begin{document}

\author{Frederic Lacombe, Stefano Zapperi and Hans J. Herrmann}
\title{Dilatancy and friction in sheared granular media}
\address{PMMH-ESPCI, 10 rue Vauquelin, 75231 Paris Cedex 05, France}
\maketitle

\begin{abstract} 
We introduce a simple model to describe the  frictional properties of
granular media under shear. We model the friction
force in terms  of the horizontal velocity $\dot{x}$ and the
vertical position $z$  of the slider, interpreting $z$ as a
 constitutive
variable characterizing the contact.  
Dilatancy is shown to play an essential role in the dynamics,
inducing a stick-slip instability at low velocity. We compute
the phase diagram, analyze numerically the model for a wide range
of parameters and compare our results with experiments on dry and wet
granular media, obtaining a good agreement. In particular, we reproduce
the hysteretic velocity dependence of the frictional force.
\end{abstract}

\section{Introduction}   \label{sec:int}
Problems related to interfacial friction
are very important from a practical and conceptual  point of view
\cite{Persson1}, and in  spite of its wide domain of
application sliding friction is still not completely  understood. In
order to construct efficient machines in engineering science,
or  to understand geophysical events like earthquakes, it is  necessary
to understand several aspects of friction dynamics.          
Beside usual solid-solid contacts, the sliding interface can be
lubricated with molecular fluids or filled by a granular gauge,
and the problem becomes strongly dependent on the internal dynamics of the
material itself.  
In experiments on lubricated surfaces with thin layers of molecular fluids
the friction force depends on the thermodynamic phase
of the lubricant, which depends on its turn on the shear stress
\cite{Gee}. The friction force is thus  directly related to the
microscopic  dynamics
of the  system and a description of sliding friction cannot be
achieved without a good microscopic understanding of the problem.  
Understanding the frictional properties of granular matter turns
out to be an even harder task, since
basic problems like stress propagation in a static
packing remain largely unsolved due to the disordered nature of the
stress repartition inside the medium. Moreover, when a granular medium
is  sheared, it reorganizes modifing the geometrical disorder. 
The microscopic arrangement of the grains and their compaction have
an important effect on the friction since, in order to deform
the medium one has to overcome several geometrical constraints. 

The understanding of
sheared granular media has recently advanced thanks to  experiments
\cite{Nas1,Gem,Marone,Veje}
and numerical studies
\cite{Thompson,Tillemans,Schollman}. The response to an external shear
stress can be
characterized by the dilatancy, which measures the
modification of the compaction of the granular medium during the flow
\cite{Thompson}. We note that in both lubricated and granular
 interfaces the friction force has a dynamical
origin. Since a sheared material modifies its own internal state 
fluidizing or changing structure, a natural approach to the  problem 
is to describe phenomenologically
this change of state and to relate it 
to  the macroscopic friction force.  
As we discussed previously, a complete theoretical description of sheared 
granular media is still not available, so that the analysis should
strongly rely on experimental data.
 
Recent experiments \cite{Nas1}, focusing on the stick-slip 
instability induced by  friction in sheared granular layers, 
helped to elucidate the role of compaction and the microscopic
origin of slip events. In particular, accurate measurement of the
friction force and of the horizontal and vertical positions of the slider
\cite{Gem} have permitted to emphasize  the connections between dilatancy and
friction. The apparatus used was composed by a slider  dragged at
constant velocity by a spring whose elongation measured the applied
shear stress. The surface of the slider was roughened
in order to avoid slip at the surface of the medium and so that 
friction would crucially depend on the internal structure
of the medium. At low velocity, a stick slip instability was observed
and related to the modification in the granular compaction.

Friction of granular layers has been mainly studied in the framework of
geophysical research \cite{Marone,Rice1} using rate and
state constitutive equations \cite{Dietrich,Rice2,Rice3,Rice4,Ruina} where the
friction force is a function of an auxiliary variable describing the
state of the interface. In this approach, one assumes that the microscopic
events causing the movement of the slider are
self-averaging and neglects the fluctuations. The quantities used in
the constitutive equations
are thus mean field-like. This assumption should be valid for sliding
friction experiments on  granular materials, where the size of the grain
is much smaller than the length of the  slider, so that the variables used
in the model (velocity, displacement or friction force), are well defined
macroscopical quantities. 

The constitutive variable, related to the microscopic dynamics of
the system, describes the dynamical history of the interface. In the case
of solid-solid interfaces \cite{Caroli} this variable was associated with
the age of the contacts and described two opposite effects: 
the age of the contact increases the static
friction force and the displacement of the 
slider renews the interface  continuously so that the friction
force decreases with velocity.
Lubricated systems have been approached similarly using the rate
of fluidization as a constitutive
variable \cite{Batista1,Persson2} which captures
two different effects. On one hand the  confinement of the thin fluid layer
induces  a glassy transition resulting in a large static friction
force. On the other hand an applied shear stress increases the
temperature of the medium, favoring fluidization, thus reducing 
the friction force, which
crucially depends on the ratio between the strength of the  two effects.

In the case of  granular media, 
a parameter suitable to  characterize the frictional behavior is
the compaction of the layers or  the height of the slider which
can be measured experimentally. Also in this case
 we can identify the competition 
between two opposite effects: 
the velocity of the slider keeps the layer
dilated, lowering the friction force,  and the weight of the
slider induces recompaction.
In this paper we present a model which includes these two effects in the
framework of rate and state constitutive equations to describe typical
effects like the stick-slip instability or the force-velocity hysteretic
loop.             

In Sec.~\ref{sec:2}  we concentrate on the description of the model, in
Sec.~\ref{sec:3}
we  describe the main results obtained by numerical  integration of
the model, in Sec.~\ref{sec:4} we present a stability analysis and the phase
diagram. Finally, Sec.~\ref{sec:5}  presents a discussion and a
summary of our results.

\section{The model}    \label{sec:2}

Here we write rate and state constitutive equations in
order to describe the frictional properties of granular media.
The dynamics of the sliding plate is described by
two constitutive equations. The first one is simply the equation of
motion for the slider block driven by a spring of stiffness $k$ and
submitted to a frictional force, which depends on velocity and
dilatancy. The second equation is the evolution law for an auxiliary
variable characterizing  the  dilatancy, which we will identify with 
the vertical position of the slider. This model could in principle
be applied to geophysical situations, although in that case
instead of a single elastic constant $k$, strain is mediated via the 
material bulk elasticity. 

The  frictional properties of  a granular medium depend
explicitly on its density: a dense granular medium submitted to a
tangential stress tends to dilate, {\it i.e } to
modify the granular packing and thus the friction force.
It is not simple to measure granular density especially for 
non homogeneous systems, but global changes  can
be  characterized by the vertical position of the sliding
plate, which is thus an excellent candidate to describe the 
state of the system. Therefore, in agreement with Ref.~\cite{Gem},  
we write the  equation of motion for the slider block as 
\begin{equation}
m \ddot{x} = k(Vt-x) - F( z,\dot{x}),
\end{equation}
where $m$ is the mass of the sliding plate, $x$ its position, $k$ 
the spring constant, $V$ the drag velocity, and $F( z,\dot{x})$ the
friction force  
depending on the velocity $\dot{x}$ and on the height of the
plate $z$.

If the slider is at rest, we need to apply a
minimal constant force $ F_0 $ in order for it to move. 
When the force exceeds $F_0$, the slider moves and dilation will
occur, reducing the friction. When the
layer is fully dilated the friction force reduces to $F_d < F_0$.
We assume that the friction force is velocity dependent when  the layer is
partially   dilated ($z<z_m$), and becomes independent on velocity  in the
stationary state, when the granular medium is fully  dilated ($z=z_m$).

In summary (in the case $\dot{x} > 0$), we write the friction force as
\begin{equation}
F( z,\dot{x}) = F_d - \beta \frac{z-z_m}{R} - \nu \dot{x}
\frac{z-z_m}{R}. 
\end{equation} 

The first two terms in Eq.~(2) give the friction force at rest
($\dot{x}=0$) as function of $z$. In the fully expanded phase ($z=z_m$), the
friction term is $F=F_d<F_0$, while in the compacted phase
$F_0=F_d+\beta z_m/R$ ($z=0$). The velocity dependence is linear, 
mediated by the factor $z-z_m $ which vanishes when the bed is fully
dilated.  These equations should be compared with those presented in
Ref.\cite{Gem}, where the second term in Eq.~(2) is not present.  

In Eq.~(2) $F(z,\dot{x})$ depends explicitly on $z$, which describes the
vertical displacement of the slider. In order to complete the
description of the 
dynamics, we must specify the evolution equation for $z$.
We write the law controlling the dilation of the granular medium
during shear as
\begin{equation}
\dot{z}=  - \frac{z}{ \eta} - \dot{x} \frac{z-z_m}{R}. 
\end{equation}

In  Eq.~(3) the second term
 dilates the support and can be seen as the response of the
granular medium to the external tangential stress: when submitted to a
shear rate $\dot{x}$, the medium dilates and $z$ increases. The factor ($ z-z_m $) reduces to zero when
the bed is fully dilated and  $z_m$ can be identified with the maximal height.

The first term allows for recompaction under the slider weight: in the case $\dot{x} = 0$ the plate falls  exponentially
fast. 
At high velocity this term will not perturb significantly the
system and the dynamics will  be stationary. We are interested in
the small velocity limit:  Eqs.~(2,3)  will display an instability below a critical drag velocity
$ V_c $, as we will show in Sec.~\ref{sec:4}.

It  is useful to rewrite the system of equations in terms of 
dimensionless variables
\begin{eqnarray*}
&& \tilde{t}=t \frac{k}{\nu}, ~~~  \tilde{\eta}=\eta \frac{k}{\nu}, ~~~
\tilde{x}= \frac{x}{R},~~~ \tilde{z}=\frac{z}{R},
~~~\tilde{z_m}=\frac{z_m}{R},\\
&&\tilde{m}= m\frac{k}{\nu^2},~\tilde{V}=V\frac{\nu}{Rk},~\tilde{v}=v\frac{\nu}{Rk},~\tilde{F_d}= \frac{F_d}{Rk},~\tilde{\beta}=\frac{\beta}{Rk}.
\end{eqnarray*}
Defining $\tilde{l}=\tilde{V}\tilde{t}-\tilde{x}-\tilde{F_d}$, 
the system becomes
\begin{eqnarray}
&& \dot{\tilde{l}}=\tilde{V}-\tilde{v}, \\
&&\tilde{ m}
\dot{\tilde{v}}=\tilde{l}+(\tilde{z}-\tilde{z_m})(\tilde{v}+
\tilde{\beta}), \\
&&\dot{\tilde{z}}=-\frac{\tilde{z}}{\tilde{\eta}}-(\tilde{z}-
 \tilde{z_m})\tilde{v}.
\end{eqnarray}
Assuming that these equation are valid for $\dot{x}>0$, 
we can analyze them for different spring constants, velocity.

\section{Numerical simulations} \label{sec:3}

We numerically solve the model (Eqs.~(4-6)) using the fourth order
Runge-Kutta method and  assuming that the slider plate sticks
when its velocity is zero. We concentrate our analysis on two sets of
parameters. The first set corresponds to experiments carried out with a
dry granular medium. We compute  the typical force-velocity diagram
in order to fix the parameters. Then using the same parameters we test the
validity of our model calculating other quantities such  as the  slider
velocity during a slip event, the spring elongation or the vertical
displacement.

A second set of parameters is used  to model wet granular
media. We recover the instability at low velocity and low spring force
and study the evolution of dilatancy and spring elongation before
reaching the steady-state.

\subsection{Dry granular media}  \label{sec:31}
 
Dry granular media exhibit
stick-slip instabilities for relatively high velocities and it
is difficult to achieve complete
vertical displacement of the slider. 
For this reason the steady sliding regime has
not been studied in detail in experiments.
In order to 
quantitatively test our model we adjust the parameters to fit the 
experimental results. We present in  Fig.~\ref{fig:f1} the force-velocity curve
during slip comparing the experimental data from
Ref.~\cite{Nas1} with the
result of the integration of the model.
The parameters used are given in the caption.
The model is able to accurately describe the first part of the
hysteretic loop (when the velocity increases), but slight deviations
appear for small velocities for which also the experimental
uncertainties are larger.

We numerically integrate the model using the previously obtained
parameters, varying the  spring constant and the driving
velocity. For slow velocity and a small spring constant the system
exhibits typical stick-slip dynamics. Fig.~\ref{fig:f2} shows the evolution of
the variables of the model in this case: the first plot (Fig.~2(a)) shows
the variation of the spring length which decreases abruptly at a regular
frequency, when the horizontal plate position increases (Fig.~2(b)).
Fig.~2(c) represents the velocity of the plate which is  followed by an
increase of the vertical position of the plate (Fig.~2(d)). We show in
Fig.~\ref{fig:f3} a more detailed study of the slider velocity 
during a slip event. Near the transition between the stick-slip and the
steady sliding regime the slider velocity appears to be almost
independent on the driving velocity, in agreement with
experiments. 
The stick-slip instability of the model is ruled by  Eqs.~(4-6) and
we present in Sec.~\ref{sec:4} the dynamical phase diagram
computed by a linear stability analysis. When the slider is
driven slowly the energy injected into the granular medium cannot keep the
layers dilated and the motion stops after a short
change in the horizontal position (slip event).

If we increase $V$ or $k$, the  energy induced by the shear is
sufficient to maintain the granular layer  dilated and 
the system evolves to a steady sliding state (cf. Fig.~\ref{fig:f4}), 
which is stable with respect  to  small perturbations. This
stationary  state, corresponds to a stable fixed point of Eqs.~(4-6)
(see Sec.~\ref{sec:4}).
If the drag velocity is very large the steady sliding state becomes
unstable due to inertial effects ($m\neq 0$) and the slider oscillates
harmonically with frequency $\omega = \sqrt{k/m}$. This effect was
experimentally observed in Ref.~\cite{Nas1}.
We have plotted the
result in Fig.~\ref{fig:f5}  for two different perturbations, 
in order to show that the amplitude of
the cycle depends on the strength of the perturbation.
  
A typical measurement performed in the framework of geophysical
research \cite{Marone,Rice1}, is the variation of the friction force with
respect to a rapid change of the driving velocity. We have simulated this
effect,  and we show the result in Fig.~\ref{fig:f6}. An increase of the
driving velocity is followed
by an increase of the friction force which then relaxes to a smaller
value.

The phase diagram corresponding to the three different
dynamical behaviors can
be calculated analytically. We present the result 
in Sec.~\ref{sec:4},
where we study the
linear stability  of the model.

\subsection{Wet granular media}    \label{sec:32}

The analysis performed in Sec.~\ref{sec:31} can be repeated for wet granular
materials. The dynamics in this case
is more stable and the stick-slip regime is
more   difficult to obtain experimentally,
since  the instability occurs at very slow velocity.
In the wet case, the presence of  water changes the dynamics of the
grains. Under shear, grains  reorganize submitted to the fluid 
viscosity but here we neglect the small hydrodynamic effects and
consider only the grain dynamics with suitable parameters.
Using the new class of parameters, we solve numerically Eqs.~(4-6) 
and identify  two regimes: steady sliding at high $V$ and 
stick-slip instability otherwise (see Sec.~\ref{sec:4} for more details).
In Fig.~\ref{fig:f8} we show a typical plot of the different quantities in the
stick-slip regime. The period of the oscillations is bigger than in the
dry case, and the fluctuations of the elongation smaller. 
One of the main difference with the dry case is the value of $\eta$
which  governs the relaxation process and which is greater in the wet
case as an effect of immersion. In Fig.~\ref{fig:f9} we show the steady state
found at high velocity. It is interesting to remark that this
behavior can be perfectly recovered with a simplified model, presented
in Ref.~\cite{Gem}, which however does not give rise to stick-slip 
instabilities. We will show in the next section that our model
is equivalent to the model of Ref.~\cite{Gem} for a given range
of parameters. Fig.~\ref{fig:f10} represents the integration in the case
$\beta > \frac{\nu R}{\eta}$ (the importance of the value of $\beta$
will be highlighted in Sec.~\ref{sec:4}).

Ref.~\cite{Gem} also reported an experiment in which the slider was
stopped abruptly, but the applied stress was not released.
Under these conditions, the medium does not recompactify towards the
initial state but remains dilated in an intermediate state.
This feature cannot be captured by our model, since the
evolution of $z$ does not explicitly depend on the 
applied stress but only on the horizontal velocity.
In order to describe this effect, we modify Eq.~(3) in order
to explicitly include a stress dependence in the evolution
of the dilatancy
\begin{equation}
\dot{z}=  - \frac{z-AF_{ext}}{ \eta} - \dot{x} \frac{z-z_m}{R}, 
\end{equation}
where $F_{ext} = k(Vt-x)$ is the applied force and $A$ is a constant.
The behavior of this model is similar to the simpler model
introduced in Section II, but the zero velocity fixed
point explicitly depend on the applied stress (i.e. $z^*=AF_{ext}$).
Fig.~\ref{fig:frelax} shows the solution of the model compared whith the
experiment of Ref.~\cite{Gem}.

\section{Linear stability}   \label{sec:4}

The simple form of Eqs.~(4-6) allows us to study
analytically the linear stability of the system.
We first concentrate on the inertial case and describe the main
results about the dynamics of our problem (fixed point, critical curve).
Next we discuss the origin of the instability and the
connections with other models. Finally we investigate the nature of 
the bifurcation.
  
\subsection{Inertial case}

All the numerical results presented above have been obtained including
inertial effects. The system of Eqs.(4-6) has a simple fixed point
\begin{equation} 
l_c=z_m \frac{V \nu+\beta}{Rk+\eta V k},
~
v_c=V,
~
z_c=z_m \frac{\eta V}{R+\eta V}.
\end{equation}
We see that $z_c$ tends to $z_m $ when $V $ tends to infinity, in
agreement with experimental result. The critical line can also be computed
explicitely in the framework of linear stability analysis. We skip the 
details of the calculations and just give the result
\begin{equation} \label{eq:i}
k^* = -\frac{ z_m \nu^2 \eta R^2 -z_m \nu \eta^2 R \beta +m R^3 \nu -m
  R^2 \eta \beta + 2 m R^2 \eta V \nu -2 m R \eta^2 V \beta +m R
  \eta^2 V^2 \nu -V^2 m \eta^3 \beta      }{\nu \eta^2 R^2 (R+\eta V)}.
\end{equation}
 Fig.~\ref{fig:f12} and Fig.~\ref{fig:f11} show the phase diagram  in
the $ k, V$ plane for the parameters used previously (in Sec.~\ref{sec:31}
and Sec.~\ref{sec:32}). For both  dry and wet granular layers we recover
the stick-slip
regime at sufficiently small $k$ and $V$. In the dry case the critical
velocity is higher than in the wet case
and we can also identify the inertial regime
on the right hand side of the phase diagram (see Fig.~\ref{fig:f12}). 
 
\subsection{Non inertial case}

If we are interested only in  low velocity displacements,  the
dynamical bifurcation line can be easily computed neglecting the
mass of the slider
\begin{equation}
k^* = (\frac{\beta}{R} - \frac{\nu}{\eta})(\frac{z_m}{R+\eta V}).
\end{equation}
Also in this case the dynamics is unstable for $k$ below the critical
line, but there is no inertial regime.
We have no experimental results to compare with this relation
which links all the relevant parameter of the model.       

Due to the simplicity of the non inertial case,
we can write our system in the traditional  form
of a Hopf bifurcation \cite{Guck}, and calculate the coefficient 
which determines the nature of the transition. 
Without the inertial term this coefficient simply reduces to
zero and therefore we have no information about the nature (super or
subcritical) of the
transition  without pushing the caluclation to higher orders or including
inertia. However, the calculation is particularly complex so we 
only analyze the problem numerically (see Sec.~IVD).

\subsection{Dynamical friction force}

The stick-slip instability is due to the dependence of the friction
coefficient on the velocity. Here we compute the
friction force corresponding to the fixed point and show that 
the sign of $ \beta / R -\nu / \eta$ plays an important role
to determine the presence of an instability. In the steady state the
friction force is given by
\begin{equation}
F_c=F_d +z_m \frac{1}{R+\eta V} (\beta + \nu V).
\end{equation}
For sufficiently high $V$, $F_c$ does not depend on $V$, in agreement
with experiments, but for relatively small velocities
$F_c(V)$ depends on $V$. The first derivative of the force is
\begin{equation}
\frac{dF_c(V)}{dV} = -\frac{R \eta}{R+\eta V}( \frac{\beta
  }{R}-\frac{\nu}{\eta} ).
\end{equation}
We can thus identify three cases:

$\bullet$  if  ($\beta /R -\nu / \eta$) is
positive then  there is a $k$ verifying Eq.~(10) and below this
$k$ the system is unstable (the derivative in $V$ of $F_c(V)$ is negative
$i.e~  F_c(V)$ decreases with $V$ ).

$\bullet$ if ($\beta /R -\nu / \eta$) is negative
the system is always stable ($k$ cannot be negative). 

$\bullet$ if ($\beta /R  - \nu / \eta =0$) then $F_c(V)$
does not depend on $V$. In this case the system is stable and we
can write the friction force as
\begin{equation} \label{eq:e}
F( z,\dot{x})= F_s +\nu \dot{z},
\end{equation} 
with $  F_s= F_d + \nu  z_m / \eta  $.
The form given in Eq.~(\ref{eq:e}) for the friction force, together with
Eq.~(3), implies a friction coefficient independent on $V$ and
a stable steady state for all the values of the parameters. 
In the limit $\eta \gg 1$, and assuming Eq.~(\ref{eq:e}),  
$\beta  $ tends to 0 and 
the dilatancy rate is given by
\begin{equation}
\dot{z}=  -\dot{x} \frac{z-z_m}{R}, 
\end{equation}
which reproduces  the model of Ref.~\cite{Gem}.

\subsection{Nature of the bifurcation line}

The calculation in the non inertial case does not allow us to know
the exact nature of the transition. Thus we investigate this 
problem numerically:
the system is perturbed near its fixed point in the vertical
position with different displacements
($0.4 \mu m - 0.002 \mu m$). 
Two final states can be obtained depending on the position in
the phase diagram: the system can evolve to the
steady state or be driven to the stick-slip cycle. In an intermediate
zone depending on the strength of the perturbation, the system can
recover both the fixed point or the stick-slip regime. 

We identify three regimes, the first  corresponds to the stick-slip regime
where,  independently
on the amplitude of the  
perturbation the system falls into a periodic cycle. In the
second regime, associated with  high driving  velocity,  the system
evolves to the stable fixed point. In the  third intermediate
regime, the final state depends on the initial perturbation: if the
perturbation is sufficiently large  the system falls into a periodic
regime, while if the perturbation is weak it evolves towards the fixed
point.   The  transition between the two  regimes
is discontinuous  ($i.e$ subcritical). Fig.~\ref{fig:f7}
shows the amplitude of the oscillations as a function of the driving 
velocity. It would be interesting to check experimentally the hysteretic
nature of the bifurcation line.
 
The presence of an hysteretic transition line could be related
to an underlying first-order phase transition in the layer density
induced by the applied stress. Recently \cite{Ber}, analyzing the results of 
photoelastic disks in a two dimensional shear cell,
it has been argued that the
density of a the granular packing would be the order
parameter of a second order phase transition induced by shear.
It would be interesting to relate the different experimental
phase transitions through a suitable microscopic 
model.

\section{Discussion and open problems}  \label{sec:5}

We have introduced a model to describe the friction force  of a sheared
granular medium, treating explicitly the dilatancy
during the slip, in the framework of rate
and state constitutive equations. This approach allows us to include in
the description the effect of the movement of the grains and the
dependence of the friction
coefficient on the dynamics of the layer. The variables used
are mean-field like, since they represent
macroscopic quantities like the position or the velocity but they are
sufficient to describe phenomenologically the system.
We have integrated the model for two sets of parameters, in order to
make quantitative predictions for two different experimental
configurations  corresponding to dry and wet granular media.

The results  are in good agreement with experiments.
In particular, we recover the hysteretic dependence of the
friction force on velocity and  obtain a good fit to the experimental
data recorded in dry granular media.
The effect of the weight of the slider plate is included in the model
and allows us to
recover a stick-slip instability at low velocity.  The physical
origin of the instability is then directly related to the recompaction of the material
under normal stress.
The dynamical phase diagram is  calculated analytically
both in the inertial and non inertial cases and inertia 
is found to change only the
high velocity part of this diagram. 
The equations used to model  the dependence of the friction law
on the external parameters include explicitly the effect of 
recompaction  in the
evolution of the vertical slider position.

The use of constitutive equations to model the friction force on
complex interfaces is the simplest way to obtain quantitative results
on the dynamics of the system. This approach provides  good results in
various fields, from  geophysics or  to nanotribology. 
In order  to include the dynamics (or thermodynamics in the case of
lubrication) of the material in the description, we need detailed informations
about the material used.
Our knowledge of sheared  granular media is very poor due to the particulate
and disordered nature of such materials and it is difficult to
characterize the internal stress and strain rate.
A precise description of the friction force for granular systems should
include some information about the stress repartition inside the
sheared material. This is a difficult problem which  even for the
simple case of  a static pile cannot be solved  completely.
In the dynamical regime,  the velocity depends on the precise nature of the
contacts and on the friction force induced by them.
Statistical models are needed to obtain 
a more complete macroscopic description based on the microscopic
grain dynamics. In this respect, the analogy with phase transitions
could be extremely fruitful.

Experiments on granular flow  over a rough inclined plane display
an interesting behavior \cite{Pouliquen,Daerr}, which is ruled by frictional properties. The
dynamic stops abruptly when the drag force decreases and the system
freezes with  grains remaining in a static configuration. 
These phenomena can be related to  the dependence of the friction force on
the velocity of the grains: an increase of the friction force when
the velocity of the layer decreases can produce an instability as in
the system discussed here. It will be interesting to see if the
methods   discussed in this paper can be applied to this and other
situations. 

We thank J. S. Rice, and S. Roux for useful discussions and
encouragements. We are grateful to J-C. Geminard for providing us with
the data of his experiments and for interesting remarks.
S. Z. is supported by EC TMR Research Network under contract
ERBFMRXCT960062.

\begin{twocolumn}


\begin{figure}   
\epsfig{file=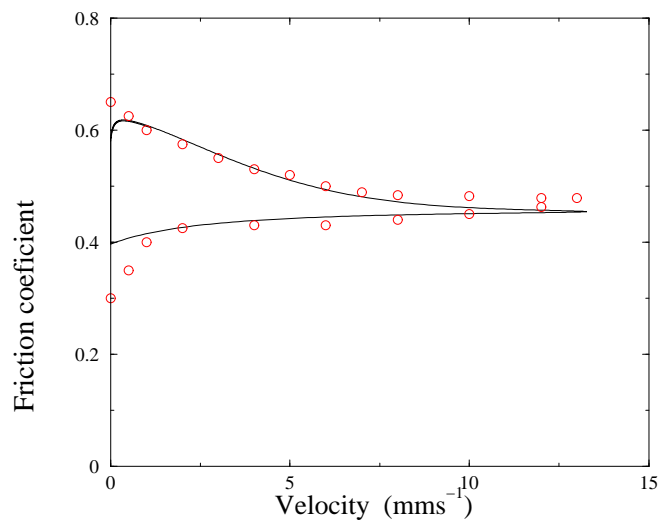,width=\linewidth} 
\caption{ Friction coefficient as a function of the slider velocity. The circles
  are the experimental data from Ref.~[3] the line is the result of the
  model with parameters: $\nu = 90Kgs^{-1}$, $\eta = 0.1 s$, $ m = 11.3 g$, $R = 8 \mu m$, $z_m = 8 \mu m$, $\beta =
  3 \frac{\nu R}{\eta}$, $k = 135 N m^{-1}$. } \label{fig:f1}
\end{figure}

\begin{figure}        
\begin{center}
\epsfig{file=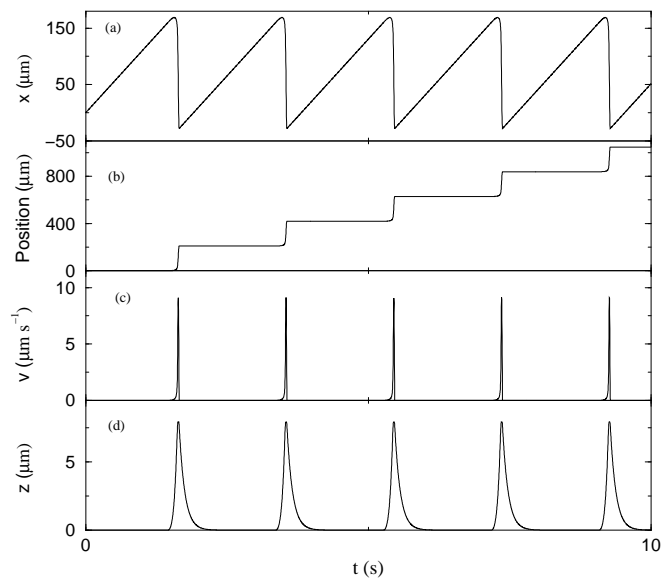,width=\linewidth,angle=0}
\caption{ Numerical calculation for dry granular  media in the
  stick-slip regime. ($\nu = 90Kgs^{-1}$, $\eta = 0.1 s$, $m = 10 g$, $R = 8 \mu m$, $z_m = 8 \mu m$, $\beta =
  3 \frac{\nu R}{\eta}$, $V = 110 \mu m s^{-1}$, $k = 135  N m^{-1}$). (a) spring elongation, (b) horizontal displacement, (c)~ slider velocity, (d)~vertical displacement .   } \label{fig:f2}
\end{center}
\end{figure}

\begin{figure}
\epsfig{file=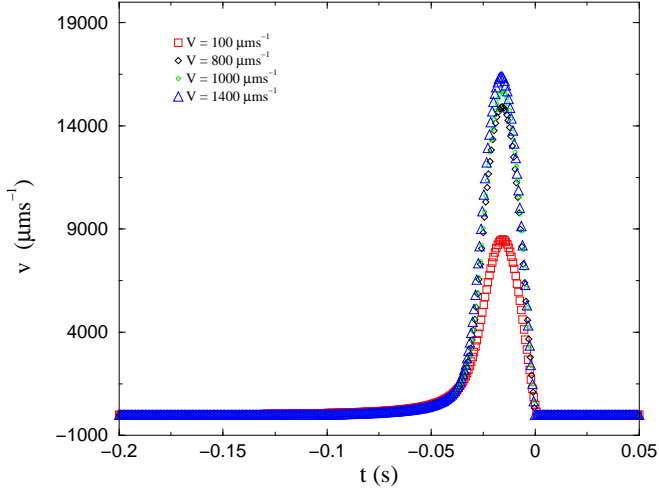,width=\linewidth} 
\caption{ Slider velocity for different loading velocities during a slip
  event. ($\nu = 90Kgs^{-1}$, $\eta = 0.1 s$, $m = 10 g$, $R = 8 \mu m$, $z_m = 8 \mu m$, $\beta =
  3 \frac{\nu R}{\eta}$, $k = 135 N m^{-1}$). Close to the transition
  point the maximum velocity is weakly dependent on the driving velocity.}\label{fig:f3}
\end{figure}

\begin{figure}
\begin{center}
\epsfig{file=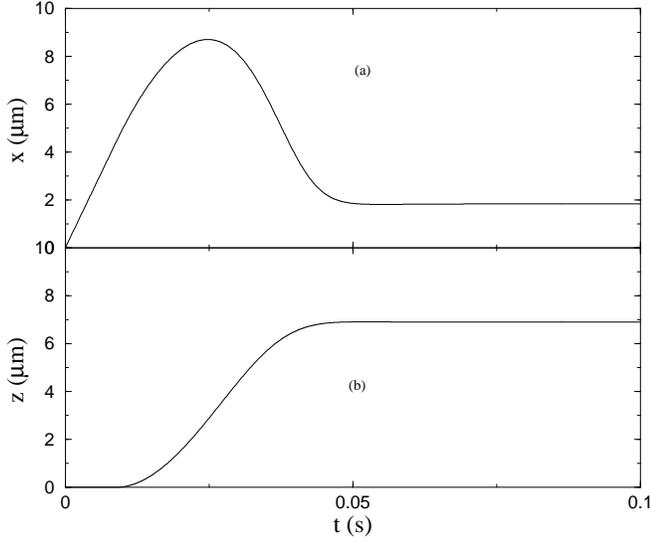,width=\linewidth}
\caption{ Evolution to 
  steady sliding state; numerical calculation for dry granular
  media. ($\nu = 90Kgs^{-1}$, $\eta = 0.1 s$, $m = 10 g$, $R = 8 \mu m$, $z_m = 8 \mu m$, $\beta =
  3 \frac{\nu R}{\eta}$, $V = 500 \mu m s^{-1}$, $k = 5 K N
  m^{-1}$). (a) spring elongation, (b) vertical position.}\label{fig:f4}
\end{center}
\end{figure}

\begin{figure}
\epsfig{file=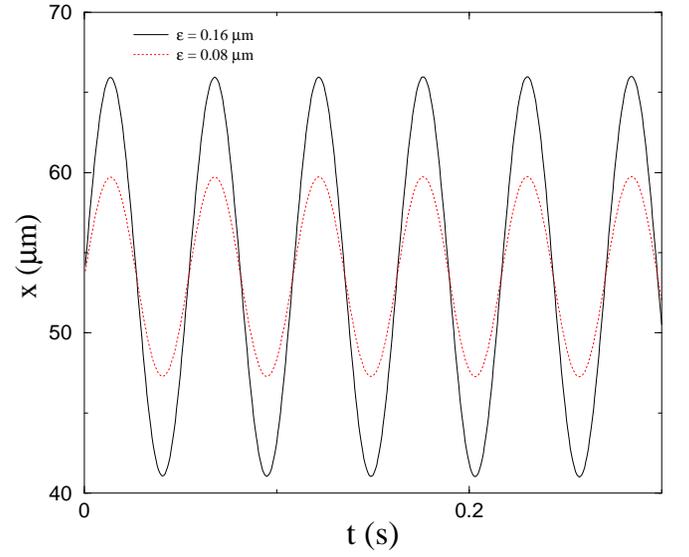,width=\linewidth} 
\caption{ Inertial oscillations. ($\nu = 90Kgs^{-1}$, $\eta = 0.1 s$,
  $m = 10 g$, $R = 8 \mu m$, $z_m = 8 \mu m$, $\beta =
  3 \frac{\nu R}{\eta}$, $V = 50 m m s^{-1}$, $k = 135 N m^{-1}$). The
  amplitude of the cycle depends on the initial condition. Here $x_0 =
  x_c,~ v_0 = v_c, ~ z_0 = z_c- \epsilon$, where $\epsilon $  is
  given in the legend, $(x_c,v_c,z_c)$ is the fixed
  point.}  \label{fig:f5}
\end{figure}

\begin{figure} 
\epsfig{file=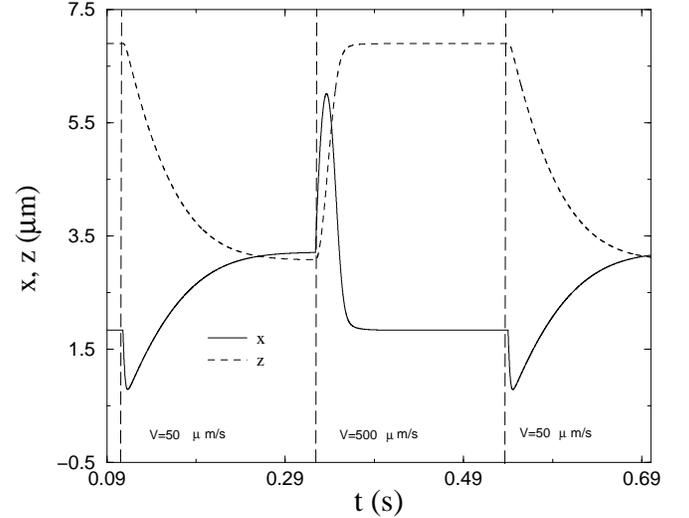,width=\linewidth} 
\caption{The system is sheared at different constant
  velocities. We show the effect of the velocity change on the spring
  elongation and on dilatancy, see Ref.~[15-17]. ($\nu =
  90Kgs^{-1}$, $\eta = 0.1 s$, $m = 10 g$, $R = 8 \mu m$, $z_m = 8 \mu m$, $\beta =
  3 \frac{\nu R}{\eta}$, $V = 50 ~~ 500 \mu m s^{-1}$, $k = 5 K N m^{-1}$).      } \label{fig:f6}
\end{figure}

\begin{figure} 
\begin{center}
\epsfig{file=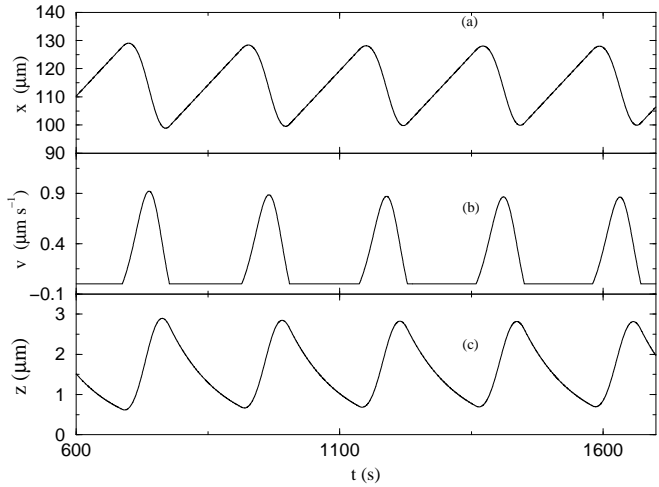,width=\linewidth,angle=0}

\caption{ Numerical calculation for wet granular  media in the
  stick-slip regime. ($\nu = 10000Kgs^{-1}$, $\eta = 100 s$, $m = 10
  g$, $R = 100 \mu m$, $z_m = 10 \mu m$, $\beta =
  15 \frac{\nu R}{\eta}$, $V = 0.2 \mu m s^{-1}$, $k = 110  N m^{-1}$). (a) spring elongation, (b) plate velocity, (c) vertical position.  }\label{fig:f8}
\end{center}
\end{figure}

\begin{figure}
\epsfig{file=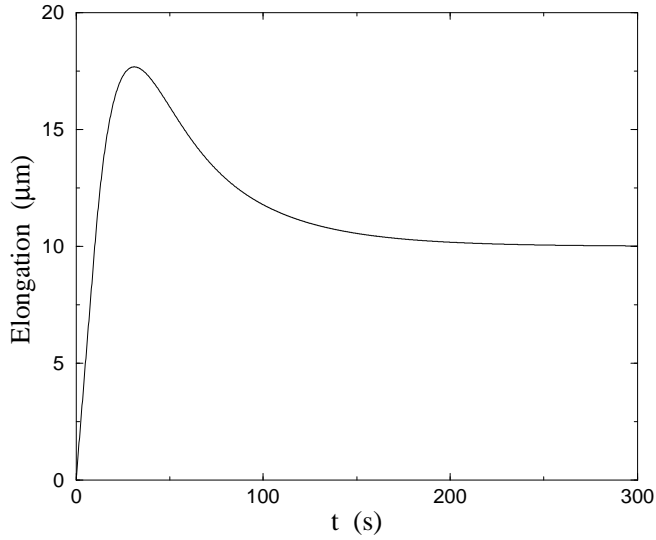,width=\linewidth,angle=0}
\caption{Steady sliding regime, for the wet case.   $\beta = \frac{\nu R}{\eta}$. ($\nu
  = 10000Kgs^{-1}$, $\eta = 100 s$, $m = 10 g$, $R = 100 \mu m$, $z_m = 10 \mu m$, $\beta =
   \frac{\nu R}{\eta}$, $k = 100  N m^{-1}$, $V = 1 \mu m s^{-1}$).}\label{fig:f9}
\end{figure}

\begin{figure}
\begin{center}
\epsfig{file=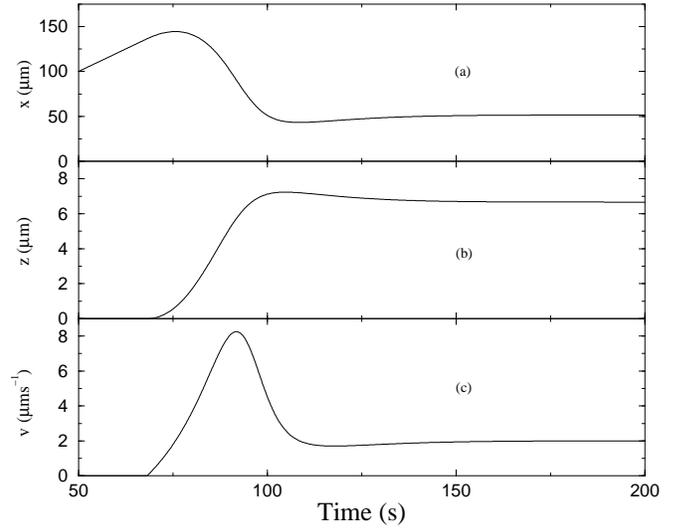,width=\linewidth,angle=0}

\caption{Steady sliding regime, effect due to $\beta > \frac{\nu
    R}{\eta}  $. ($\nu = 10000Kgs^{-1}$, $\eta = 100 s$, $m = 10 g$, $R = 100 \mu m$, $z_m = 10 \mu m$, $\beta =
  15 \frac{\nu R}{\eta}$, $V = 2 \mu m s^{-1}$, $k = 110  N m^{-1} $). (a) spring
  elongation, (b) horizontal position, (c) plate velocity. With
 $ \beta =0$ and $ \eta \gg 1$ we recover the result of Ref.[4].    }\label{fig:f10}
\end{center}
\end{figure}

\begin{figure}
\begin{center}
\epsfig{file=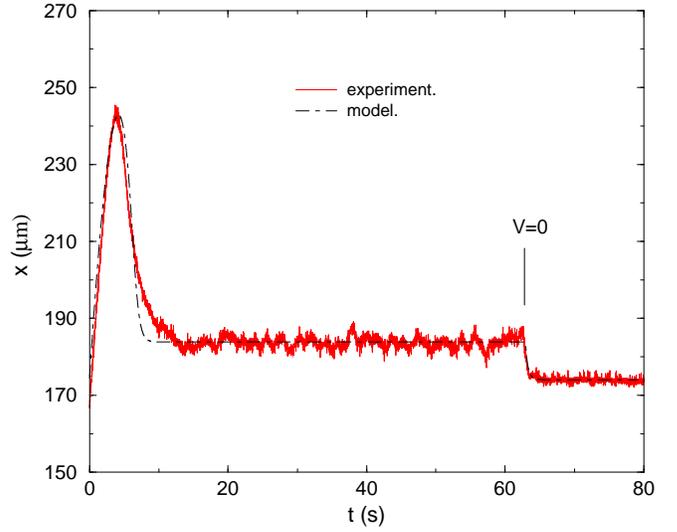,width=\linewidth,angle=0}

\caption{The plate is stopped in the steady sliding regime
(at time $t \approx 60s$) keeping the applied stress constant. We compare the
solution of the model with the experimental data of Ref.[4]. ($\nu = 10000Kgs^{-1}$, $\eta = 30 s$, $m = 14.5 g$, $R = 59 \mu m$, $z_m = 6 \mu m$, $\beta =
  0.01 \frac{\nu R}{\eta}$, $V = 28.35 \mu m s^{-1}$, $k = 189.5  N
  m^{-1} $, $A=0.137 \mu mN^{-1}$).   }\label{fig:frelax}
\end{center}
\end{figure}

\begin{figure}
\epsfig{file=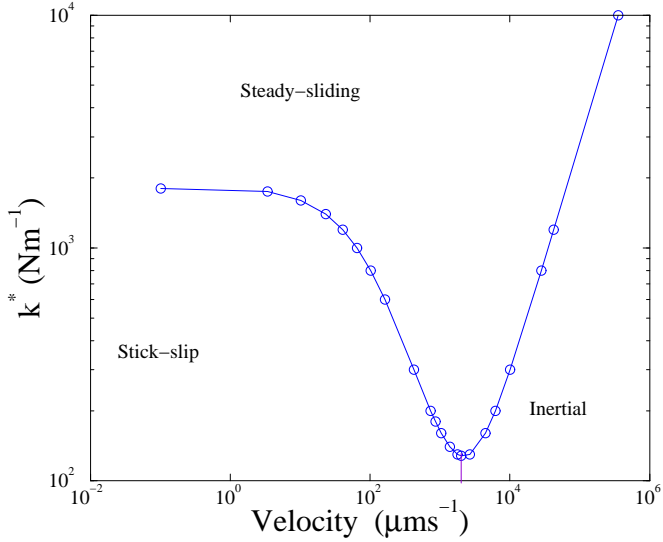,width=\linewidth} 
\caption{Dynamical phase diagram. Parameters here have been chosen in
  order to  reproduce the data of Ref.[3]. on dry granular
  media, the circles are the results of numerical calculation. ($\nu = 90Kgs^{-1}$, $\eta = 0.1 s$, $m = 11.3 g$, $R = 8 \mu m$, $z_m = 8 \mu m$, $\beta =
  3 \frac{\nu R}{\eta}$).  We can  identify three different zones. At
  low velocity and low $k$ the system exhibits a stick-slip
  instability, while at high velocity we observe an inertial
  regime. We also observe an intermediate region characterized by steady
  sliding.} \label{fig:f12}
\end{figure}

\begin{figure}
\epsfig{file=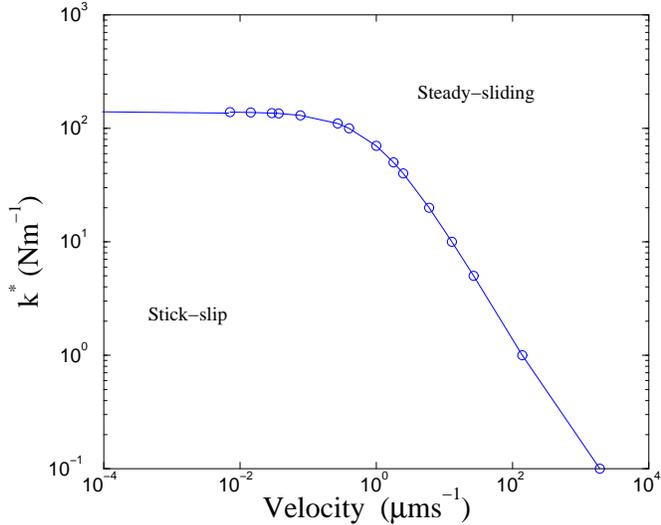,width=\linewidth } 
\caption{Dynamical phase diagram for parameters corresponding to the
  wet case. At high $V$ and $k$ we have a steady-sliding phase which
  becomes unstable at $k=k^*$. ($\nu =
10000Kgs^{-1}$, $\eta = 100 s$, $m = 10 g$, $R = 100 \mu m$, $z_m = 10
  \mu m$, $\beta = 15 \frac{\nu R}{\eta}$). The circles represent numerical
  calculation.}\label{fig:f11}
\end{figure}

\begin{figure} 
\epsfig{file=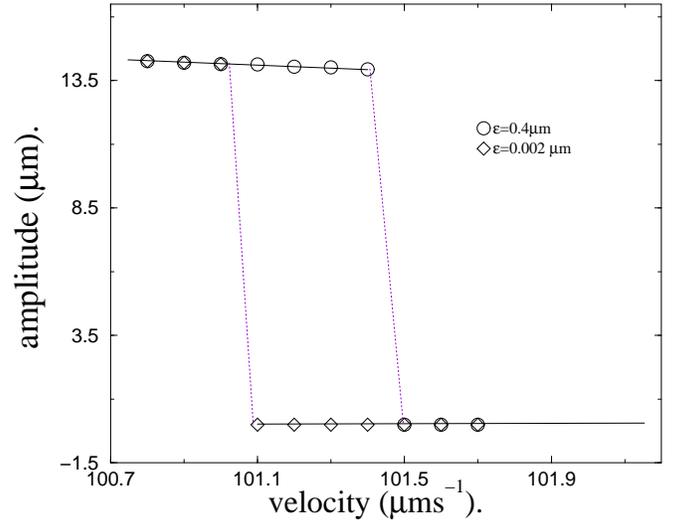,width=\linewidth} 
\caption{Amplitude of the oscillation as a function of driving
  velocity. We perturb the system
  near  its fixed point in the vertical direction. Two
  amplitudes are used  in order to identify
  the three regimes, the diagram represents the final state of the
  system as a function of the perturbation $\epsilon$. ($\nu =
  90Kgs^{-1}$, $\eta = 0.1 s$, $m = 10 g$, $R = 8 \mu m$, $z_m = 8 \mu m$, $\beta =
  3 \frac{\nu R}{\eta}$, $k = 800 N m^{-1}$).} \label{fig:f7}
\end{figure}

\end{twocolumn}

\end{document}